\documentclass{aa} 
\usepackage{natbib}
\bibpunct{(}{)}{;}{a}{}{,} % to follow the A&A style
\usepackage{graphicx}
\usepackage{hyperref}
\usepackage{txfonts}
\newcommand{\beq}{\begin{equation}}
\newcommand{\eeq}{\end{equation}}
\def\alp{\mbox{$\alpha$}}

\def\farcm{\hbox{$.\mkern-4mu^\prime$}}

\def\arcmin{\hbox{$^\prime$}}
\def\arcsec{\hbox{$^{\prime\prime}$}}
\def\solar{\mbox{$_{\normalsize\odot}$}}

\def\micron{\mbox{$\mu$m}}
\newcommand{\AmS}{{\protect\the\textfont2
  A\kern-.1667em\lower.5ex\hbox{M}\kern-.125emS}}
\newcommand{\lsim}{\ \raise
-2.truept\hbox{\rlap{\hbox{$\sim$}}\raise5.truept\hbox{$<$}\ }}
\newcommand{\gsim}{\ \raise
-2.truept\hbox{\rlap{\hbox{$\sim$}}\raise5.truept\hbox{$>$}\ }}
\newcommand{\simsim}{\ \raise
-2.truept\hbox{\rlap{\hbox{$\sim$}}\raise5.truept\hbox{$\sim$}\ }}

\begin{document}

\title{Recent star formation at low metallicities. The star-forming region NGC~346/N66 
in the Small Magellanic Cloud from near-infrared VLT/ISAAC observations\thanks{Based 
on observations made with ESO Telescopes at the La Silla Paranal Observatory under 
program ID 063.I-0329.} \fnmsep \thanks{Table~\ref{t:15stars} is available in its entirety 
only in electronic form at the CDS via anonymous ftp to {\tt \url{ftp://cdsarc.u-strasbg.fr}} 
or via {\tt \url{http://cdsweb.u-strasbg.fr/cgi-bin/qcat?J/A+A/515/A56}}
}}

%\subtitle{Probing Recent Star Formation 
%in Low-Metallicities\thanks{Based on observations made with ESO
%Telescopes at the La Silla Paranal Observatory under program ID
%063.I-0329}}

%   \subtitle{Photospheric line asymmetries and wavelength shifts}

\author{Dimitrios A. Gouliermis \inst{1}
\and 
Joachim M. Bestenlehner \inst{1,2}
\and 
Wolfgang Brandner \inst{1} 
\and 
Thomas Henning \inst{1}
}
%\and 
%M\'onica Rubio \inst{3}
%\and 
%Rodolfo H. Barb\'a \inst{4}
%}

\offprints{D.\ A.\ G. \email{dgoulier@mpia.de}}

\institute{Max-Planck-Institut f\"{u}r Astronomie, K\"{o}nigstuhl 17,
69117 Heidelberg, Germany\\ 
%\email{\small dgoulier@mpia-hd.mpg.de, brandner@mpia-hd.mpg.de, henning@mpia-hd.mpg.de}
\and
Armagh Observatory, College Hill, Armagh BT61 9DG, UK
}
%\and 
%Departamento de Astronom\'{\i}a, Universidad de Chile, Casilla 36-D,  
%Santiago, Chile
%\and 
%Departamento de F\'{\i}sica, Universidad de La Serena, Benavente 980, 
%La Serena, Chile
%}
        
   \date{Received ... ; accepted ...}

% \abstract{}{}{}{}{} 
% 5 {} token are mandatory

  \abstract
{The emission nebula N66 is the brightest H~II region in the Small Magellanic Cloud (SMC), 
the stellar association NGC~346 being located at its center. The youthfulness of the region 
NGC~346/N66 is well documented by studies of the gas and dust emission, and the detection 
in the optical of a rich sample of pre-main sequence (PMS) stars, and in the mid- and far-IR of 
young stellar objects (YSOs). However, a comprehensive study of this region has not been 
performed in the near-IR that would bridge the gap between previous surveys.}
{We perform a photometric analysis on deep, seeing-limited near-IR VLT images of the region 
NGC~346/N66 and a nearby control field of the SMC to locate the centers of active high- and 
intermediate-mass star formation by identifying near-IR bright objects as candidate stellar 
sources under formation.}
{We use archival imaging data obtained with the high-resolution camera ISAAC at VLT of 
NGC~346/N66 to construct the near-IR color-magnitude (CMD) and color-color diagrams 
(C-CD) of all detected sources.  We investigate the nature of all stellar populations in the 
observed CMDs, and we identify all stellar sources that show significant near-IR excess 
emission in the observed C-CD. We, thus, select the most likely young stellar sources.} 
{Based on their near-IR colors, we select 263 candidate young stellar sources. This sample 
comprises a variety of objects, such as intermediate-mass PMS and Herbig Ae/Be stars and 
possible massive YSOs, providing original near-IR colors for each of them. The spatial 
distribution of the selected candidate sources indicates that they are located along the 
dusty filamentary structures of N66 seen in mid- and far-IR dust emission and agrees 
very well with that of previously detected candidate YSOs and PMS stars.}
{Our study provides an original accurate set of near-IR colors for candidate young stellar 
sources. This provides significant information about the star formation process in 
NGC~346/N66, but does not establish the types of these objects, which requires the 
construction of complete spectral energy distributions for individual sources from 
multiwavelength data. This would be an important follow-up study to that presented here.}


\titlerunning{ISAAC imaging of the SMC star-forming region NGC~346/N66}
\authorrunning{D. A. Gouliermis et al.}
\maketitle

%%%%%%%%%%%%%%%%%%%%%%%%%%%%%%%%%%%%%%%%%%%%%%%%
\begin{figure*}[t!]
\centering 
\includegraphics[width=1.0\textwidth]{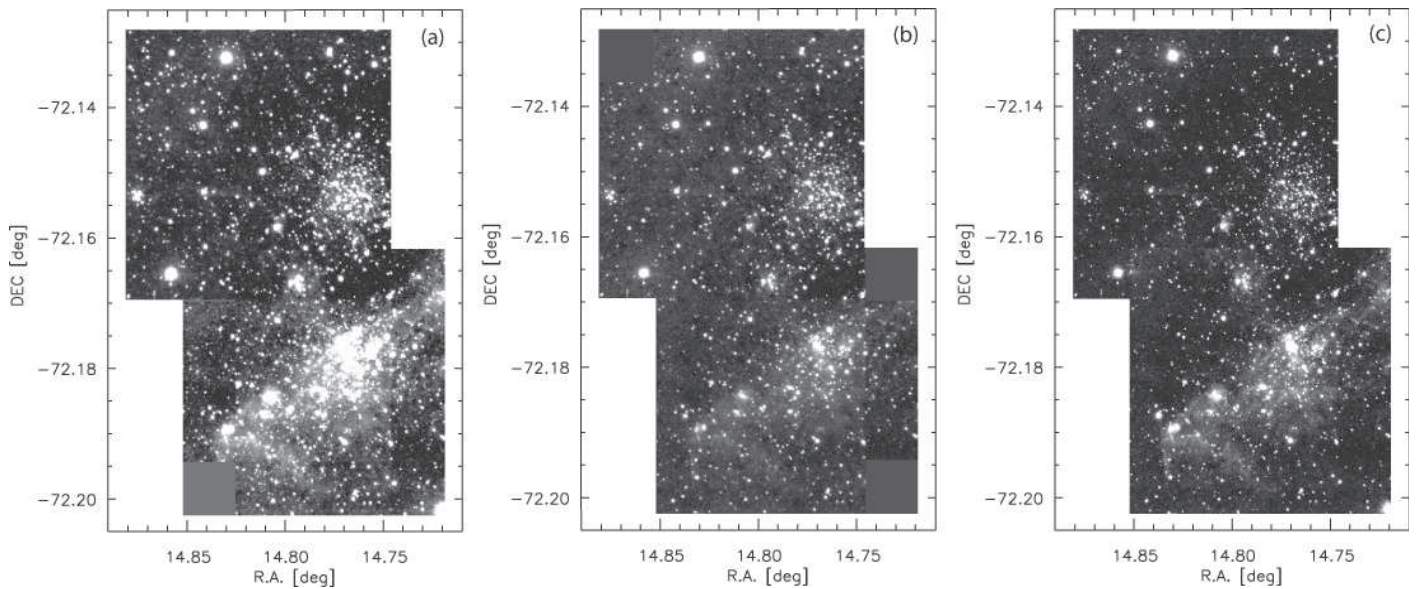} 
\caption{Mosaic images from the combination of the northern and center
jittered ISAAC frames of the area of NGC~346/N~66 in the (a) $J$-, 
(b) $H$-, and (c) $K_{\rm s}$-band.} 
\label{f:jitimages} 
\end{figure*} 
%%%%%%%%%%%%%%%%%%%%%%%%%%%%%%%%%%%%%%%%%%%%%%%%

\section{Introduction}

Located in the stellar constellation {\sl Tucana}, the Small Magellanic Cloud (SMC), is 
an excellent laboratory for investigating the star formation processes and the associated 
chemical evolution of dwarf galaxies. Its present subsolar chemical abundance ($Z=0.004$; 
$\sim$ 20\% of solar) implies that this galaxy may have characteristics similar to those in 
earlier times  of the evolution of the universe. Since SMC is so close to our Galaxy, it is 
therefore an excellent laboratory among the large collection of dwarf irregulars and blue 
compact galaxies for the study of resolved extragalactic stellar populations and star-forming 
regions. The young stellar association NGC~346 (RA (J2000)~$=$~00$^{\rm h}$~59$^{\rm m}$~18$^{\rm s}$, 
DEC (J2000)~$=$~$-$72$^\circ$~10$'$~48$''$) is a large star-forming cluster in the SMC, 
located at a distance of about 60.6~kpc from us \citep{hilditch2005}. It is  embedded in the 
brightest {\sc H~ii} region of the SMC, which is referred to as LHA~115$-$N66 or N66 
\citep{henize1956}. With an H$\alpha$ luminosity almost 60 times higher than the star-forming 
region of Orion \citep{kennicutt84}, N66 has a diameter of about $7'$ corresponding to 
approximately 123 pc. 

The star-forming region NGC~346/N66 comprises a variety of young stellar populations 
\citep{gouliermis06}. The stellar association NGC~346, located at the center of the so-called
nebular `bar' of N66, hosts the largest sample of O-type stars in the entire SMC \citep{massey1989, 
walborn2000, evans2006}. Studies based on deep imaging with the {\sl Hubble Space Telescope} 
show that the vicinity of the whole region of NGC~346/N66 is also very rich in low- and 
intermediate-mass PMS stars, some of which exhibit recent star formation with an age \lsim~5~Myr, 
while others belong to an older underlying population of age \lsim~10~Myr \citep[see e.g.,][]{hennekemper08}. 
Additional evidence of the youthfulness of this region comes from observations with the {\sl Spitzer 
Space Telescope} and the discovery of 111 candidate massive young stellar objects (YSOs) with 
2~\lsim~$M$/M\solar~\lsim~17 \citep{simon07}. Nevertheless, although these objects should emit in 
near-IR bands, no detailed study in such wavelengths exists in the literature. 

Pioneering work on the dust and gas content of NGC~346/N66 was performed by \citet{rubio00} 
and \citet{contursi00}, who found a correlation between H$_{2}$ 
infrared emission and CO lines, characteristic of a photo-dissociated region (PDR).
A PDR develops when the far-UV radiation of the bright OB stars reaches the 
surface of the parental molecular cloud. The degree of ionization decreases
outwards, and a thin barrier develops that segregates the ionized from the 
atomic gas. While  H$_{2}$ is not fully ionized behind this front but partly 
dissociated, CO molecules located little deeper in the cloud are more easily 
dissociated by absorbing UV photons. \citet{rubio00} and \citet{contursi00} 
inferred that star formation in NGC~346/N66 
has taken place as a sequential process in the ``bar'' of N~66, which these authors
define as the oblique bright emission region extending from southeast to 
northwest centered on NGC~346. This process results in several embedded 
sources, seen as IR-emission peaks in the 2.14~\micron\ H$_{2}$ line and 
the ISOCAM LW2 band (5 - 8 \micron). These peaks are alphabetically 
numbered from ``A'' to ``I'' \citep{rubio00, contursi00}, the association 
NGC~346 itself coinciding with peak ``C''.

%%%%%%%%%%%%%%%%%%%%%%%%%%%%%%%%%%%%%%%%%%%%%%%%
\begin{figure*}[t!]
\centering 
\includegraphics[width=1.0\textwidth]{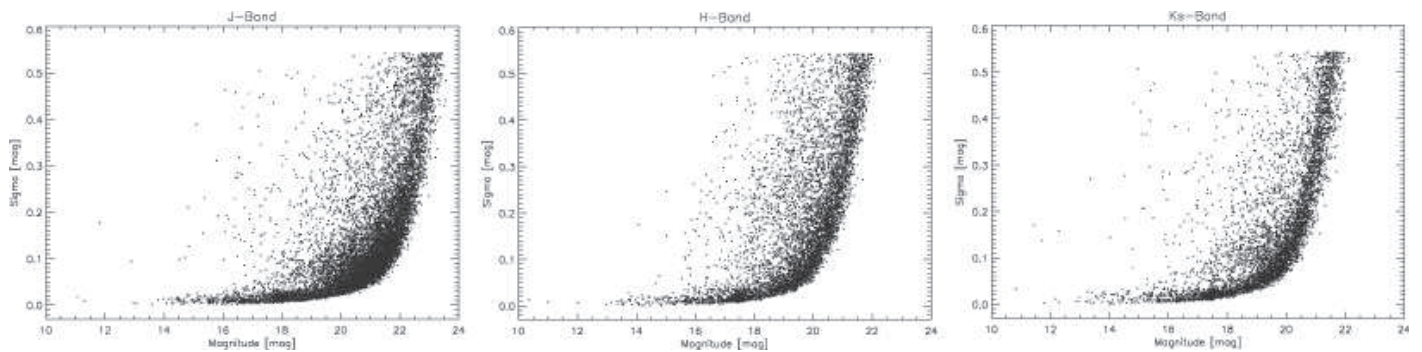} 
\caption{Typical photometric uncertainties in the three $J$, 
$H$, and $K_{\rm s}$ bands, derived by DAOPHOT for all detected 
sources.} 
\label{f:photerr} 
\end{figure*} 
%%%%%%%%%%%%%%%%%%%%%%%%%%%%%%%%%%%%%%%%%%%%%%%%

Studies with the {\sl Advanced Camera for Surveys} (ACS) on-board the {\sl Hubble Space 
Telescope} (HST) have shown that the PMS stellar content of NGC~346/N66 covers a mass 
range in the subsolar regime. These studies suggest that recent star formation occurred around 
3 - 5 Myr ago \citep{nota2006, sabbi07} there being an underlying older PMS population,
which indicates that there were earlier star formation events that occurred about 10~Myr ago 
\citep{hennekemper08}. The PMS population is mainly centrally concentrated apart from the 
association NGC~346 in a number of subclusters with clustering properties quite similar to those 
of Milky Way star-forming regions \citep{schmeja09}. Furthermore, the intermediate-age 
star cluster BS~90 \citep{bica95} with an age of about 4.5~Gyr \citep{rochau07} is also located 
in this part of the SMC, projected in front of N66.  

This photometric study of the star-forming region NGC~346/N66 focuses on 
identifying young stellar sources, mainly intermediate- and high-mass
PMS stars of the region, stars that have not yet started their lives on the main 
sequence. The PMS phase in the evolution of stars with masses up
to $\sim$~6~M{\solar} corresponds to the time between the gravitational
core collapse, which forms the protostar (on the birthline), and the ignition
of hydrogen in the formed star, placing it on the Zero-Age Main Sequence 
(ZAMS). During this evolutionary phase, the observed radiation from the 
star is affected significantly by circumstellar disks of dust and gas, 
formed by matter infalling during the collapse of the rotating core, and 
surface activity. Examples of these typical PMS stars, the {\sl T~Tauri} stars, 
thus, exhibit prominent optical emission lines, which are
understood to stem from chromospheric heating, periodic fluctuations in
light that indicate rotating star-spots, variability, and excess broadband
flux in UV and IR, and are sometimes associated with molecular
outflows, winds, or accretion \citep[e.g.,][]{lada91}. Intermediate-mass 
(2~\lsim~$M$/M{\solar}~\lsim~10) PMS stars are called Herbig AeBe 
(HAeBe) stars \citep[e.g.,][]{perez97}. Being more massive 
analogues of T~Tauri stars, they are PMS A- and B-type stars that exhibit  
emission lines produced by both strong stellar winds and the cocoons of 
remnant gas from which they collapsed. They typically contain circumstellar 
disks and therefore have spectral energy distributions (SEDs) of young 
stellar objects (YSOs) of class II. They range in age between 0.5 and 5 Myr, 
similar to T~Tauri stars. Because of the aforementioned
characteristics, PMS stars are visible in near-IR wavebands. 

While extensive observations have been performed of young stellar sources 
in the region NGC~346/N66 at optical \citep{nota2006, gouliermis06} and 
mid- and far-IR \citep{bolatto07, simon07} wavelengths, there has been no 
comparative investigation in the near-IR. The present study aims to 
close this gap in the available spectral coverage for this region by acquiring 
near-IR data to characterize with greater accuracy its young stellar population. 
This will also aid our understanding of star formation in the low-metallicity environment 
of the SMC. In this paper, we present our ground-based near-IR 
photometry derived from observations with VLT/ISAAC of the region NGC~346/N66. 
We present the observational material used and its reduction in Sect.~2, and we discuss the
photometric process in Sect.~3. The various observed stellar populations and the 
corresponding stellar systems comprised in the observed field are discussed 
in terms of variations in the constructed color-magnitude diagrams in Sect.~4. We
present the constructed near-IR color-color diagram of all detected sources 
and apply a selection criterion by identifying stars currently 
forming by means of their near-IR excess emission in terms
of their positions in this diagram in Sect.~5. In the same section, the spatial 
distribution and the nature of the selected young stellar sources is also 
discussed. Finally, conclusive remarks on this study are given in Sect.~6.

%%%%%%%%%%%%%%%%%%%%%%%%%%%%%%%%%%%%%%%%%%%%%%%%
\begin{figure}[b!]
\centering 
\includegraphics[width=0.975\columnwidth]{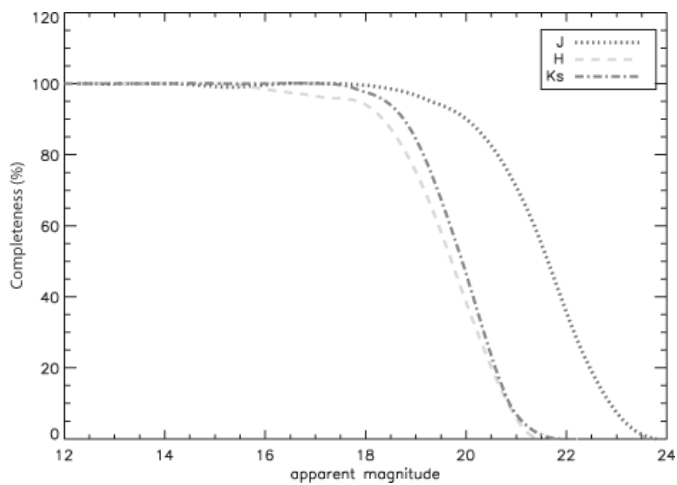} 
\caption{Completeness for all stars detected in the three 
observed fields in all three $J$, $H$ and $K_{\rm s}$ filters.} 
\label{f:completeness} 
\end{figure} 
%%%%%%%%%%%%%%%%%%%%%%%%%%%%%%%%%%%%%%%%%%%%%%%%

\section{Observations and data reduction}

Red giants, stars close the end of their life usually have 
high mass-loss rates, while stars at the earliest stages of their 
formation are embedded into dense molecular clumps and cores. In both 
cases, the dust shells or disks around the objects of interest absorb almost all 
of the visible radiation, which is re-radiated at longer wavelengths. As a 
consequence, absorption decreases very rapidly with increasing 
wavelength \citep[e.g.,][]{joyce92,glass99}, the extinction coefficient 
at 2.2~$\mu$m being approximately 10\% of that at 500~nm. 
Here we are interested in young stellar sources associated 
with circumstellar dusty shells or disks, characterized by bright 
IR excess emission.

\subsection{Observations}

The near-IR images of NGC~346/N66 are obtained 
within the ESO Program ID 63.I-0329 (PI: M. Rubio) with 
the {\sl Infrared Spectrometer And Array Camera} (ISAAC), mounted on 
the Nasmyth-Focus B of UT1 ({\sl Antu}) at the Very Large Telescope 
(VLT). The images were taken between 24 and 26 Sep. 1999,
under fair seeing conditions (FWHM between 0.6\arcsec\ and 1.0\arcsec).
They were used partly to detect YSOs in the 
region of N66 \citep{simon07}, but have never been presented in their
complete spatial and wavelength coverage. The observations were
performed with the short wavelength arm of ISAAC, using the 
1024~$\times$~1024 HgCdTe Hawaii Rockwell array. The pixel scale of 
the Hawaii detector is 0.1484\arcsec/pixel, providing a maximum 
field-of-view of 152~$\times$~152~arcsec$^2$. The images were 
obtained in the filters $J$ (1.25 $\mu$m), $H$ (1.65 $\mu$m), and 
$K_{\rm s}$ (2.16 $\mu$m). 

At wavelengths longer than $\sim$~2.2~$\mu$m,  
the thermal background is dominated by atmospheric and telescope emission, 
leading to a highly variable sky brightness in the infrared. Detector 
cosmetics and instabilities also illustrate the need for an accurate sky subtraction. 
This is achieved by mean of the technique of 
{\sl jittering}\footnote{\url{http://www.eso.org/projects/dfs/papers/jitter99/}}, 
which is available for the short-wavelength and some
long-wavelength imaging templates of ISAAC.  A set of 10 to 100 frames are 
combined to form one final ``jittered'' frame. At the beginning and end of each 
night, twilight flats are taken for each filter. The
following morning, dark images of the detector are taken for all 
{\sl detector integration times} (DITs) used during the previous night. 
The area of NGC~346 was divided into two jittered frames, each consisting 
of 10 single frames. An additional jittered frame was taken for a control field of the galaxy,
located to the south of the system. For each single frame, the total integration
time was 60s, taken in 5s (DIT) exposures of 12 sub-integrations (NDIT). 
The field-of-view is about 2.5$^\prime$~$\times$~2.5$^\prime$, corresponding to about
44~$\times$~44~pc$^2$ at the SMC \citep[distance from us 60.6 kpc;][]{hilditch2005}.
We retrieved the dataset of these observations, including the science frames, 
twilight flats, dark images, and standard stars from the ESO Science Archive
Facility\footnote{\url{http://archive.eso.org/cms/}}.

\subsection{Data reduction}

To reduce the data, we used 
ESO's ISAAC pipeline, which is based entirely on the libraries of the 
data reduction package ECLIPSE\footnote{\url{http://www.eso.org/eclipse}}. 
The package consists of the main eclipse-library and the additional instrument
specific pipeline packages. Another algorithm
of ECLIPSE is the {\sl jitter} routine, which implements efficient 
filtering and processing methods for infrared data reduction. The detector 
pixels have intrinsically different sensitivities because of the quantum efficiency 
variations from pixel to pixel. There is also a readout gradient of the array, 
which itself is not homogeneously illuminated by the telescope. 
These effects, as well as artifacts caused by dust on the optical surface, 
are corrected within the data reduction process. 

The dark current of the ISAAC Hawaii array is low, 
so the detector bias,  also called the ``zero level offset'', 
is the dominant feature of the dark frames.  A {\sl master dark}, 
constructed with ECLIPSE, was subtracted from the science
frames. The effects of dust, the inhomogeneity of the
illumination, and pixel to pixel variations were then removed 
using a {\sl master flat-field} frame. Since no flat-field screen 
exists at UT1, for ISAAC only twilight flats are available. For the 
subtraction of both sky and bias, the darkest twilight
flat frame was used. After this subtraction, the remaining 15 twilight
flats were combined to produce a master flat-field with ECLIPSE, which also 
creates a bad pixel map output file. The {\sl jitter} routine
in ECLIPSE was then used to reduce the science data for the frames 
observed with acceptable seeing, by filtering out low-frequency sky
variations from the set of jittered images, a method called {\sl
sky combination}.  Two sets of jittered images were combined for the whole area of NGC~346.
The corresponding final frames are shown for each filter in Fig.~\ref{f:jitimages}. 
The association NGC~346, the numerous young compact star clusters located in the 
bright {\sc Hii}-region N~66, and the  intermediate-age star cluster BS~90 are 
easily identified in these images. The photometry, discussed in the following 
section, was performed on the single jittered frames for each filter. 
The derived catalogs were then matched to produce the complete photometric 
catalog of the observed regions.

%%%%%%%%%%%%%%%%%%%%%%%%%%%%%%%%%%%%%%%%%%%%%%%%
\begin{figure}[t!]
\centering
\includegraphics[width=0.975\columnwidth]{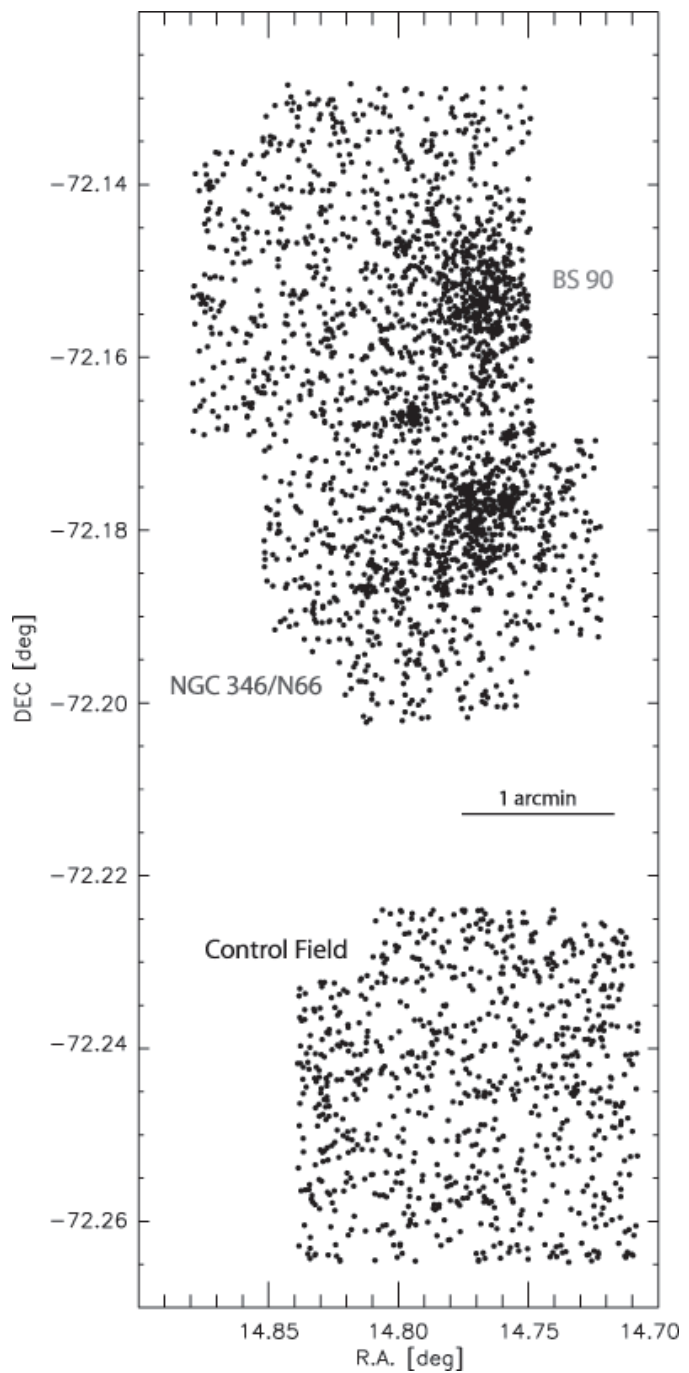} 
\caption{Chart of all stars detected with good photometric accuracy in all three 
filters in all three fields observed with VLT/ISAAC. This map shows the 
relative positions of the observed fields: The northern field that includes the 
intermediate-age cluster BS~90, the central field, where the association NGC~346 and the 
main part of the H~II region N66 are located, and the remote control-field, which  
covers the representative stellar populations of the general SMC field in the area.
North is up and East is to the left of the map.}
\label{f:xymap}
\end{figure}
%%%%%%%%%%%%%%%%%%%%%%%%%%%%%%%%%%%%%%%%%%%%%%%%

%For the reduction and analysis of astronomical data the NOAO package 
%was developed by the National Optical Astronomy Observatories\footnote{\url{http://www.noao.edu/}} for
%IRAF. DAOPHOT is a part of the DIGIPHOT package within NOAO for 
%obtaining precise photometric indices and astrometric
%positions for stellar objects in crowded fields in two dimensional
%digital images. 

%Since the overlap between neighboring stars leads to large
%uncertainties in the aperture photometry, the obtained magnitudes are
%not accurate. Another uncertainty is due to the ``blurring'' of the
%stars, caused by the spreading of the light of the point sources.  This
%blurring is basically the result of the quality of the imaging system
%and the seeing conditions (twinkling of the light due to atmospheric
%turbulence) at the time of observation. This effect can be quantified by
%the so-called {\sl Point Spread Function} (PSF), which represents the
%image of a typical point source for these observations. 

\begin{table*}[]
\caption{Sample of the photometric catalog of stars found in this study 
in the region of NGC~346/N66 and a nearby control field in all three 
$J$, $H$, and $K_{\rm s}$ filters with VLT ISAAC imaging.
This table is available in its entirety at the CDS. \label{t:15stars} }
\centering
\begin{tabular}{r c c r r c c c c c c}
\hline\hline
\multicolumn{1}{c}{ID} & R.A. & Decl. & \multicolumn{1}{c}{X} & \multicolumn{1}{c}{Y} & $J$ & $\sigma_{J}$ & $H$ & $\sigma_{H}$ & $K_{\rm s}$ & $\sigma_{K_{\rm s}}$\\
 & (deg) & (deg) & \multicolumn{1}{c}{(px)} & \multicolumn{1}{c}{(px)} & (mag) & (mag) & (mag) & (mag) & (mag) & (mag)\\
\hline
1	&	14.79218	&	$-$72.16765	&	531.765	&	155.835	&	16.855	&	0.013	&	16.853	&	0.012	&	16.920	&	0.023	\\
2	&	14.79437	&	$-$72.16741	&	548.261	&	150.011	&	16.011	&	0.014	&	15.968	&	0.013	&	15.979	&	0.034	\\
3	&	14.85005	&	$-$72.16258	&	965.954	&	31.829	&	17.483	&	0.007	&	17.098	&	0.010	&	17.097	&	0.018	\\
4	&	14.79519	&	$-$72.16639	&	554.363	&	124.946	&	17.957	&	0.029	&	18.110	&	0.054	&	18.113	&	0.043	\\
5	&	14.79583	&	$-$72.16637	&	559.162	&	124.505	&	15.665	&	0.004	&	15.664	&	0.009	&	15.731	&	0.018	\\
6	&	14.79584	&	$-$72.16676	&	559.283	&	134.180	&	18.141	&	0.025	&	17.840	&	0.037	&	17.714	&	0.055	\\
7	&	14.80519	&	$-$72.16749	&	629.400	&	151.956	&	18.724	&	0.023	&	18.307	&	0.026	&	18.074	&	0.027	\\
8	&	14.81704	&	$-$72.16776	&	718.270	&	158.700	&	19.061	&	0.017	&	18.626	&	0.031	&	18.561	&	0.033	\\
9	&	14.84279	&	$-$72.16213	&	911.515	&	20.758	&	17.483	&	0.006	&	17.117	&	0.010	&	17.097	&	0.016	\\
10	&	14.85175	&	$-$72.16175	&	978.702	&	11.646	&	19.181	&	0.021	&	19.364	&	0.065	&	19.252	&	0.059	\\
11	&	14.85197	&	$-$72.16337	&	980.320	&	51.276	&	18.321	&	0.010	&	17.905	&	0.022	&	17.907	&	0.023	\\
12	&	14.78242	&	$-$72.16232	&	458.576	&	25.405	&	17.969	&	0.014	&	17.820	&	0.019	&	17.878	&	0.023	\\
13	&	14.78620	&	$-$72.16421	&	486.962	&	71.662	&	16.760	&	0.008	&	16.605	&	0.009	&	16.645	&	0.011	\\
14	&	14.79341	&	$-$72.16464	&	541.046	&	82.180	&	18.041	&	0.008	&	17.567	&	0.016	&	17.488	&	0.014	\\
15	&	14.80191	&	$-$72.16452	&	604.786	&	79.288	&	18.263	&	0.011	&	17.814	&	0.019	&	17.679	&	0.017	\\
\hline
\end{tabular}
%\begin{list}{}{}
%\item[] (X,Y) coordinates are given in respect to the central field (0,0) 
%point. Table \ref{t:15stars} is published in its entirety in the 
%electronic edition of the {\sl Astronomy \& Astrophysics Journal}. 
%A portion of the complete catalog is shown here for guidance regarding 
%its form and content.
%\end{list}
\end{table*}

\section{Photometry}

%\subsection{Photometric Process}\label{s:photproc}

Our photometry was performed within the Image Reduction and Analysis 
Facility (IRAF) system\footnote{\url{http://iraf.noao.edu/}}, with the 
package DAOPHOT\footnote{\url{http://iraf.noao.edu/scripts/irafhelp?daophot}} 
\citep{stetson87}. The effect of both the quality of the imaging system and 
the seeing conditions on the obtained magnitudes of the stars is 
quantified by the {\sl point spread function} (PSF), which represents 
the image of a typical point source in the observations.
DAOPHOT provides accurate PSF photometry of crowded stellar
fields. The magnitudes of all stars detected in the observed jittered images are
defined relative to the brightness of {\sl Vega}.
We first have to specify the input parameters of DAOPHOT to optimize 
the detection and photometry of stars, before using the routine {\tt daofind} to 
detect all stellar sources in the science frames. The routine 
{\tt phot} is then used to perform
aperture photometry of the detected stars and determine their 
instrumental magnitudes. The radius of the aperture is selected 
to be between 1 and 2 times that of the FWHM, depending on the effective 
seeing.  For individual stars, the aperture radius can be up to 5 or more of
that of the FWHM. 

%%%%%%%%%%%%%%%%%%%%%%%%%%%%%%%%%%%%%%%%%%%%%%%%
\begin{figure}[b!] 
\centering 
\includegraphics[width=0.975\columnwidth]{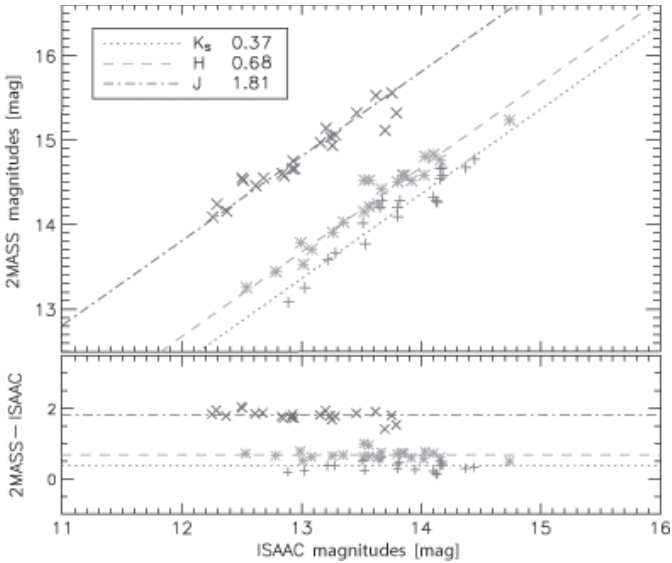} 
\caption{Comparison of our photometry with that derived by 2MASS
for more than 20 common stars, located in all three observed fields
for all three $J$, $H$, $K_{\rm s}$ filters. {\sl Top}: ISAAC versus (vs.) 2MASS 
photometry for the common stars. The linear fits that define the 
offset for each filter are shown with different lines and colors. The 
corresponding offsets applied to the absolute calibration of our 
photometry are also given for every filter. {\sl Bottom}: Dispersion in the 
2MASS$-$ISAAC magnitudes for the common stars as a function of 
ISAAC magnitudes. The offsets for each filter are again defined by 
different lines. The standard deviation in the points derived from 
the corresponding mean values is the measurement of 
the uncertainties in our photometric  calibration for each filter 
(see Sect.~\ref{calib}).} 
\label{f:2masscalib} 
\end{figure} 
%%%%%%%%%%%%%%%%%%%%%%%%%%%%%%%%%%%%%%%%%%%%%%%%

The typical PSF of our images was constructed by interactively selecting 
approximately 20 isolated bright stars per frame using  DAOPHOT 
{\tt pstselect}. The PSF was modelled to be the 
sum of an analytic bivariate Gaussian function and empirical corrections
from the best Gaussian of the true observed brightness values within 
the average profile of several stars in the image. This process and the computation of 
the PSF to be fitted was performed by the routine  {\tt psf}. PSF photometry was 
performed with the routine {\tt allstar}, which 
after classifying the stars in groups, compiles a catalog of the 
most likely candidate stars, based on
their PSF fitting and the physical conditions, and subtracts them from
the original image. The photometric process was repeated for the
subtracted frames and the magnitudes of its newly discovered stars were
determined. Approximately 30\% more stars were found in the region of
NGC~346 from the second photometry run, but most of them have
large photometric uncertainties. The final numbers of identified stars in each
filter is 11~900 in $J$, 6~406 in $H$, and 5~837 in $K_{\rm s}$. We match 
these photometric catalogs with a procedure developed in IDL 
to identify stars in common. After selecting the stars with the highest photometric 
quality, i.e., stars with photometric uncertainties equal or smaller than 0.1~mag, 
we identified 2~783 stars in both the $J$ and $H$ filters, 3~067 stars in $J$ 
and $K_{\rm s}$, 2~550 stars in $H$ and $K_{\rm s}$, and 2~506 stars in all three
filters collectively.

\subsection{Completeness and photometric accuracy}\label{s:compl}

Typical uncertainties in our photometry are shown in 
Fig.~\ref{f:photerr}. The completeness of our photometry is 
evaluated by artificial star experiments, by adding artificial 
stars into the original frames with the DAOPHOT 
subroutine {\tt addstar}. A few hundred artificial stars were
added to each observed frame for every filter. This process
was repeated several times, and the completeness of our photometry was 
derived by applying the photometric process to each new frame and 
comparing the final stellar catalogs with those of the original artificial stars based
on the coordinates of the stars. The derived completeness of our photometry 
is shown in Fig.~\ref{f:completeness} for all three bands. Based on the 
completeness measurements and the photometric uncertainties, the
limiting magnitudes of our photometry are $J$~$\simeq$~22.0~mag,
$H$~$\simeq$~20.5~mag, and $K_{\rm s}$~$\simeq$~20.5~mag. 

%%%%%%%%%%%%%%%%%%%%%%%%%%%%%%%%%%%%%%%%%%%%%%%%
\begin{figure*}[t!]
\centering 
\includegraphics[width=1.0\textwidth]{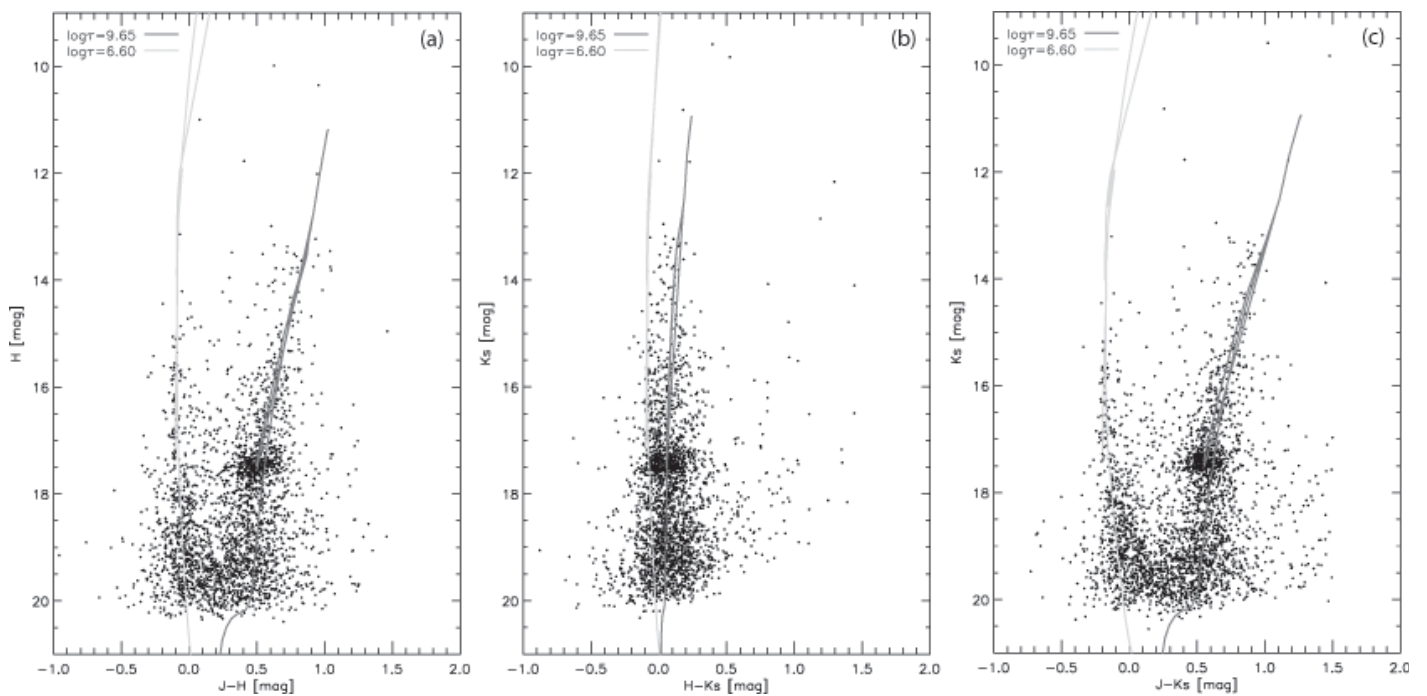} 
\caption{CMDs of all stars found in all observed fields of the 
region NGC~346/N66 and its nearby control field of the SMC. CMDs 
in different combinations of the three bands are shown. Specifically, 
(a) $J-H$, $H$, (b) $H-K_{\rm s}$, $K_{\rm s}$, and (c) $J-K_{\rm s}$, 
$K_{\rm s}$. Two indicative isochrones from the Padova grid of evolutionary 
models \citep{girardi2002}, corresponding to ages $\sim$~4~Myr and 
$\sim$~4.5~Gyr are over-plotted to demonstrate the differences 
of the comprised stellar populations.} 
\label{f:totalcmds} 
\end{figure*} 
%%%%%%%%%%%%%%%%%%%%%%%%%%%%%%%%%%%%%%%%%%%%%%%%

\subsection{Photometric calibration}\label{calib}

We calibrate the magnitudes of the stars with near-IR absolute photometry 
of the same field from the Two Micron All Sky 
Survey\footnote{\url{http://irsa.ipac.caltech.edu}} (2MASS). The 2MASS 
filter system 
agrees very well with that of ISAAC, and  
the 2MASS Point Source Catalog (All-Sky 2003) is therefore particularly suitable for the calibration of our
photometry. The application ALADIN\footnote{\url{http://aladin.u-strasbg.fr/aladin.gml}} was used
to identify all stars found with 2MASS in the same
field-of-view as ours and to compare the magnitudes
measured from both ISAAC and 2MASS data sets. To calculate the magnitude offsets 
between the two photometries, we selected more than 20 bright stars with the highest quality 
photometric data from our photometric catalog in each filter. The comparison of our photometry with that
of 2MASS is shown in Fig.~\ref{f:2masscalib}. The offsets per filter 
are calculated in terms of the median of the differences between the magnitudes of the 
stars, which are common to both ISAAC and 2MASS samples. The apparent
magnitude offsets are 1.812~mag in $J$, 0.678~mag
in $H$, and 0.367~mag in $K_{\rm s}$. The uncertainties in this calibration can be quantified
by the standard deviations derived from the dispersion in 2MASS$-$ISAAC magnitudes
for the common stars (Fig.~\ref{f:2masscalib} - bottom). They are $\sigma_{J} = 0.14$~mag, 
$\sigma_{H} = 0.12$~mag, and $\sigma_{K_{\rm s}} = 0.13$~mag, respectively.

\subsection{The photometric catalog}\label{s:photcat}

We derived celestial coordinates for all stars detected with high quality photometry
($\sigma \leq$~0.1~mag) from their pixel coordinates in the final 
jitter FITS images and used the applications {\tt xy2sky} and {\tt sky2xy} 
available from the {\sl World Coordinate Systems} (WCS) {\sl Tools}\footnote{Available
at \url{http://tdc-www.harvard.edu/wcstools/}} (ver. 3.7.2). We 
transformed the (X,Y) coordinates of the stars into celestial coordinates according 
to the astrometric corrections provided in the FITS header 
of the corresponding jittered images by applying {\tt xy2sky}. We then   
transformed the celestial coordinates of all stars with the use of {\tt sky2xy} 
into a common (X,Y) pixel coordinate system with respect to the central field, where NGC~346/N~66 
is mainly observed, by using the astrometric corrections provided in the FITS header 
of the $K_{\rm s}$ image of this field. A sample of the final compiled catalog of the stars 
found in all three filters is given in Table~\ref{t:15stars}.  This table is available in its entirety 
at the CDS. In Fig.~\ref{f:xymap}, the map of all sources detected in all three wavebands is shown. 

%This table is available in its entirety in the electronic edition of this paper.
%The complete photometric catalog will be available online.

\section{Observed stellar populations}\label{s:stelpop}

In our subsequent analysis, we consider only sources detected with the highest
quality photometric parameters and with photometric uncertainties based on 
the PSF fitting of $\sigma \leq$~0.1~mag in every waveband. The region 
NGC~346/N66 is known to host a mixture of stellar populations 
\citep[e.g.,][]{gouliermis06}, including the evolved SMC field stars, 
the young main-sequence (MS) and pre-main sequence (PMS) populations of 
the association NGC~346 and its vicinity \citep[e.g.,][]{hennekemper08},
and the $\sim$~4.5~Gyr-old faint MS and bright RGB stars of the 
cluster BS~90 \citep{rochau07}.

%%%%%%%%%%%%%%%%%%%%%%%%%%%%%%%%%%%%%%%%%%%%%%%%
\begin{figure*}[t!]
\centering 
\includegraphics[width=1.0\textwidth]{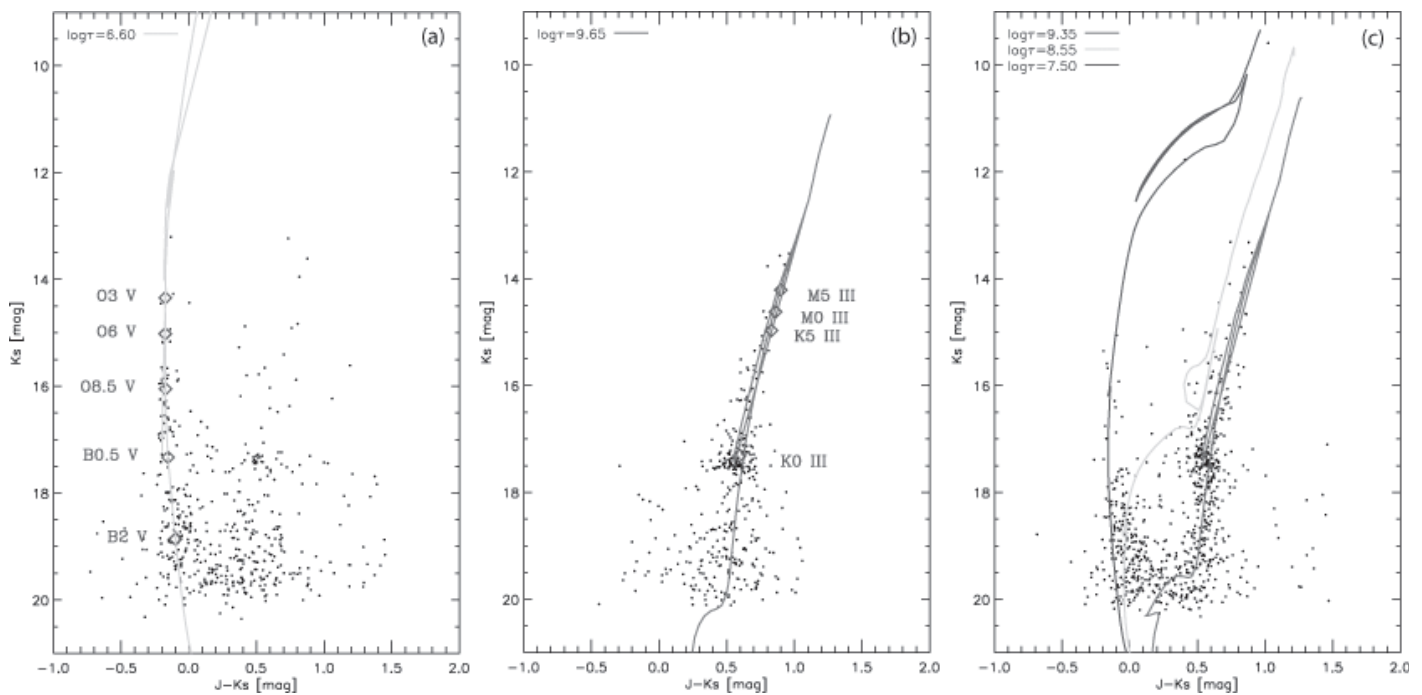} 
\caption{$J-K_{\rm s}$, $K_{\rm s}$ CMDs of stars found in two
selected parts of the observed region NGC~346/N66, for (a) a circular
area of radius $\sim$~10~pc centered on NGC~346, (b) a circular
area of radius $\sim$~8.5~pc centered on BS~90, and (c) the control
field alone. The variations in the CMD from one area to another 
are highlighted by the superimposed isochrone models and typical positions of 
early-type MS stars and late-type giants (see Sect.~\ref{cmdvar}).} 
\label{f:partscmds} 
\end{figure*} 
%%%%%%%%%%%%%%%%%%%%%%%%%%%%%%%%%%%%%%%%%%%%%%%%

\subsection{Color-magnitude diagrams}\label{s:cmds}

The color-magnitude diagrams (CMDs) were compiled from our photometry 
for all stars detected in both the area of NGC~346/N66 (where BS~90 
is also included) and the control field, specifically the $J-H$ versus (vs.) 
$H$, $H-K_{\rm s}$ vs. $K_{\rm s}$, and $J-K_{\rm s}$ vs. $K_{\rm s}$ 
CMDs, shown in Fig.~\ref{f:totalcmds}. The variety of stellar 
types in the observed region is illustrated by the superimposed 
isochrone models. These evolutionary models, which are designed for 
the ESO Imaging Survey WFI $UBVRIZ$ and SOFI $JHK$ {\sc VEGAmag} 
systems, were developed by \citet{girardi2002}. The superimposed isochrones, 
corresponding to ages $\sim$~4~Myr and $\sim$~4.5~Gyr, were 
selected to correspond to the metallicity of the SMC, namely $Z=0.004$ \citep{haser1998, 
bouret2003}. Both the bright MS of NGC~346 and the red giant branch (RGB) of 
the field and BS~90 are clearly defined in the CMDs of Fig.~\ref{f:totalcmds}, 
these features being the most clearly distinguished in the 
$J-K_{\rm s}$,~$K_{\rm s}$ CMD. To determine the interstellar 
reddening of the observed stellar populations, we assume a Galactic 
interstellar extinction law \begin{equation} R_V =  \frac{A_V}{E_{B-V}}~~, 
\label{extinction_law}\end{equation} where $R_V$ is roughly equal to 
$3.14 \pm 0.10$ \citep{schultz1975}. The Galactic extinction in 
near-IR wavebands is given by $A_J = 0.282 \times A_V$, $A_H = 0.175 
\times A_V$ and $A_{K} = 0.112 \times A_V$ \citep{rieke1985}. Assuming 
these factors, we found that  to fit the young isochrone to 
the observed bright MS population in the CMDs of Fig.~\ref{f:totalcmds}, 
an extinction of $A_V \simeq 0.3$~mag should be applied. On the other hand, 
the observed RGB stars do not appear to be affected by any significant reddening in the 
CMDs ($A_V \simeq 0.02$~mag). This implies that 
the younger MS stars are still embedded within dust, while the older 
population of the field and BS~90 are not located in regions of high 
nebulosity, i.e., BS~90 is located in front of N66, and consequently their 
light does not suffer from extinction.

\subsection{Variations in the color-magnitude diagram}\label{cmdvar}

The stellar systems NGC~346 and BS~90 represent two quite different 
types of clusters both from morphological and evolutionary 
point-of-views. Their evolutionary difference is naturally 
based on the stellar member populations of each system, which
should define the prominent features in the corresponding CMDs 
of the systems. This becomes clearer when the CMDs of selected 
areas centered on these systems, comprising the most representative 
stellar populations in the systems, are constructed. In
Fig.~\ref{f:partscmds}, the $J-K_{\rm s},~K_{\rm s}$ CMDs of 
(a) a circular area of radius $\sim$~35\arcsec\ (10~pc) centred on
NGC~346, and (b) a circular area of radius $\sim$~0.5\arcmin\ (8.5~pc) 
centred on BS~90 are shown. These radii do not correspond to the 
sizes of the systems, but are selected to be very close to their centers 
to isolate the most prominent stellar populations in each 
system. The corresponding CMD of the control field is also shown 
for comparison (Fig.~\ref{f:partscmds}c). 

The variations in the CMD from one area to the next 
are indicated by the superimposed isochrone models in 
Fig.~\ref{f:partscmds}. The stellar population of 
the star-forming association NGC~346, as seen in the CMD of 
Fig.~\ref{f:partscmds}(a), includes 
mostly young bright MS stars with a small contamination of the general 
field of the SMC, as this is defined by the CMD of the control 
field shown in Fig.~\ref{f:partscmds}(c). Typical positions of 
early-type stars are also shown in Fig.~\ref{f:partscmds}(a) based 
on the study of \citet{hanson97}. On the other hand, the stellar content 
of the intermediate-age star cluster BS~90, seen in the CMD of 
Fig.~\ref{f:partscmds}(b), is characterized 
by the prominent RGB and few supergiants of age $\sim$~4.5~Gyr. The 
positions of late-type giants are also plotted based on  
\citet{peletier89} and \citet{1998gaas}. Isochrone fitting on the 
CMDs of Fig.~\ref{f:partscmds} allows us 
to derive an indicative age for the stellar systems in the region. 
For NGC~346, we find an age $\sim$~3.9~Myr, consistent with that found by, 
e.g., \citet{nota2006}. The CMD on BS~90 is most accurately fitted by the isochrone of 
an age $\sim$~4.5~Gyr, in good agreement with the value previously found by
\citet{sabbi07} and \citet{rochau07}.   It should be noted that BS 90 is a 
populous cluster, of large spatial extent, and therefore the contamination 
of the stellar populations observed in the area of the association NGC~346
by those of BS~90 should not be  negligible. We discuss this in detail in Sect.~\ref{s:ysoselect}. 
Finally, the CMD of the 
control field provides evidence of a variety of stellar populations, which are  
the results of different star formation events, of a wide range of ages, from young 
MS populations with age $\sim$~30~Myr, an order of a 
magnitude older than those in NGC~346, to evolved giants of age \gsim\,2~Gyr.
We note that near-R photometry of early-type MS stars is quite 
insensitive to ages. Therefore, the ages given here are only indicative.

The bright main sequence stars in NGC~346/N66 are
known to be O-, B-, or early A-type stars of masses between 3 and
60 M\solar, while its low- and intermediate-mass stellar content is 
characterized by a large sample of pre-main sequence (PMS) stars 
\citep[e.g.,][]{hennekemper08}. The most massive of these stars,
such as Herbig Ae/Be stars, are possibly also located in the red part of the 
main-sequence of Fig.~\ref{f:partscmds}(a).  However, our 
photometry is not in general deep enough to detect significant numbers of low-mass PMS stars.
The part of the CMD between the MS and the RGB should also host classical 
Be stars \citep[see e.g.,][]{bik2006}. To perform a more accurate 
identification of these sources and to utilize the near-IR three-band 
color-color diagram of the detected sources in the young association NGC~346,
we focus our subsequent analysis on the central area of NGC~346/N66.

%%%%%%%%%%%%%%%%%%%%%%%%%%%%%%%%%%%%%%%%%%%%%%%%
\begin{figure*}[t!]
\centering 
\includegraphics[width=0.995\textwidth]{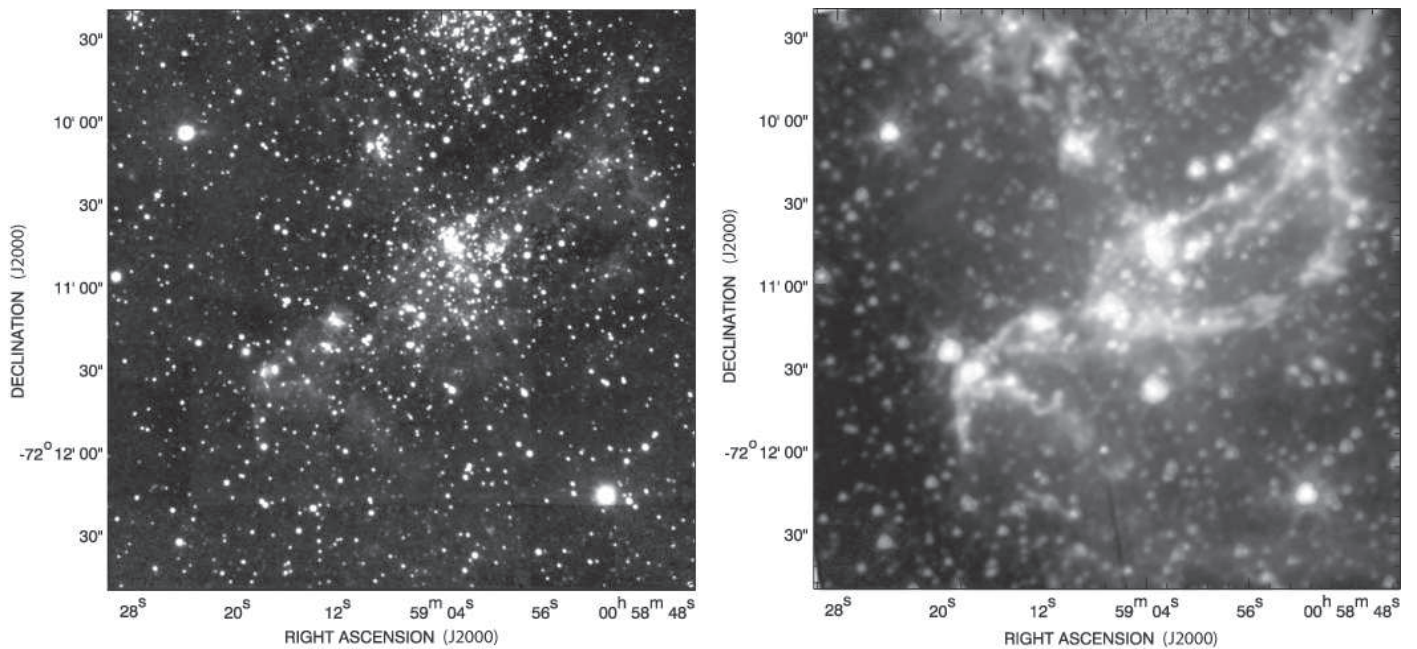} 
\caption{Color-composite images of the main part of the region 
NGC~346/N66. {\sl Left:} Image constructed with the combination 
of the jittered images in $J$ (blue), $H$ (green), and $K_{\rm s}$ 
(red) from VLT/ISAAC imaging. {\sl Right:} Image produced from 
the $K_{\rm s}$ ISAAC image (blue) and {\sl Spitzer}/IRAC 
3.6~\micron\ and 8~\micron\ filters (green and red respectively).
ISAAC images were constructed with DIMSUM (Deep Infrared Mosaicing 
Software), an external package of IRAF, especially designed
to produce accurate sky subtracted images from dithered 
observations.\label{f:colima}} 
\end{figure*} 
%%%%%%%%%%%%%%%%%%%%%%%%%%%%%%%%%%%%%%%%%%%%%%%%

\section{Stellar sources under formation}\label{s:ysopms}

There are no comparable studies to ours in the SMC, but there have been 
a few
previous near-IR photometric investigations of resolved populations in 
star-forming regions of the low-metallicity environment of the Large Magellanic 
Cloud (LMC). A combination of ground-based near-IR data and space optical 
observations of the 30 Dor Nebula by \cite{rubio98} helped identify numerous stellar 
IR sources in or near the bright nebular filaments west and northeast of R136, 
suggesting that a new stellar generation is being produced by the energetic 
activity of the massive central 
cluster in the remanent interstellar material around its periphery. 
Near-IR photometry of the second largest {\sc Hii} region in the LMC, N11B, 
also highlighted several bright IR sources \citep{barba03}. 
Several of these sources have IR colors with YSO 
characteristics, and they are prime candidates to be intermediate-mass 
Herbig Ae/Be (HAeBe) stars. 

Deep near-IR imaging of the N159/N160 star-forming region in the LMC detected 
candidate HAeBe stars down to $\sim$~3~M{\solar}, based on their near-IR 
colors and magnitudes \citep{nakajima05}. Two embedded HAeBe clusters were also 
discovered, one of them, and two neighboring {\sc Hii} regions, providing hints 
of the beginning of sequential cluster formation in N159S. The spatial distributions of the 
HAeBe and OB clusters indicate large-scale sequential cluster formation over the entire 
observed region from N160 to N159S. Near-IR photometry obtained to study the stellar content of the 
LMC star-forming region N4 is used to study the connection between the interstellar 
medium and its stellar content \citep{contursi07}. This analysis found several 
bright IR sources with characteristics of massive, early O-type stars. However, 
according to these authors, IR spectroscopy of these sources would confirm 
their very young and massive nature. 

\cite{chen09} presented the most complete identification of 
YSO candidates in the LMC {\sc Hii} complex N44. These authors combined 
mid- and far-IR {\sl Spitzer Space Telescope} imaging with complementary 
ground-based imaging in $UBVIJK$ to classify the YSOs into Types I, II, and 
III, according to their spectral energy distributions (SEDs). In their sample, $\sim$~65\% 
of these objects were resolved into multiple components or extended sources. 
The distribution of the YSO candidates compared with those of the underlying 
stellar population and interstellar gas illustrates a correlation between the current 
formation of O-type stars and previous formation of massive stars, providing evidence 
of triggered star formation in N44.

In connection to the aforementioned studies, our photometry, although not very deep, 
provides original near-IR colors of candidate young sources in the vicinity 
of N66. However, from near-IR photometry 
alone it is impossible to accurately determine the nature of the 
most IR-bright sources, unless this photometry is part of a multi-band 
optical {\sl and} IR investigation, as we discuss later. In the present study, 
we provide only a first set of accurate near-IR colors for all young objects 
in the region, and select the most interesting IR-bright objects from our 
photometric catalog, based on their near-IR colors and their near-IR 
excess inferred from the $J-H$,  $H-K_{\rm s}$ color-color diagram (C-CD).
This selection is described in the following section.

%%%%%%%%%%%%%%%%%%%%%%%%%%%%%%%%%%%%%%%%%%%%%%%%
\begin{figure*}[t!]
\centering 
\includegraphics[width=0.975\textwidth]{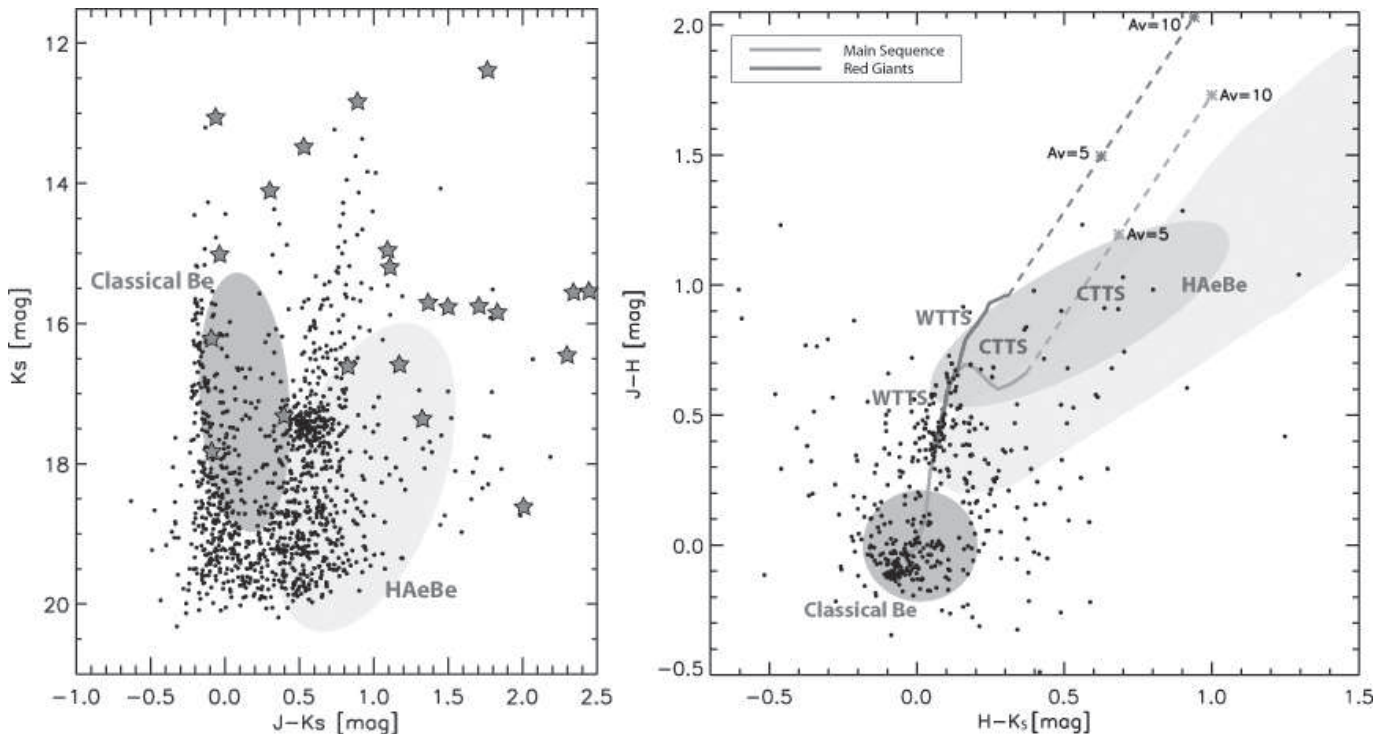} 
\caption{$J-K_{\rm s}$ vs. $K_{\rm s}$ color-magnitude diagram (CMD)
and the corresponding $J-H$ vs. $H-K_{\rm s}$ color-color diagram (C-CD)
of all sources found in the main part of the star-forming region 
NGC~346/N66 with high photometric accuracy ($\sigma$~\lsim~0.1~mag) 
from our VLT/ISAAC photometry.  Typical locations of Herbig Ae/Be stars (HAeBe; yellow 
shaded areas) and classical Be stars (red shaded areas) are shown based on the studies 
by \citet{lada92},  \citet{dougherty94}, \citet{meyer97} and \citet{eiroa02}.  While T~Tauri 
stars (TTS) are fainter than our detection limit in the CMD, typical positions of classical 
T~Tauri stars (CTTS; blue shaded area), and the general loci of weak-line T~Tauri 
stars (WTTS)  are also shown in the C-CD for completeness. In the CMD, the green 
``$\star$'' symbols represent the extinction-corrected positions of the massive YSOs 
from the samples of \citet{hanson97} and \citet{bik2006}. The green and red lines 
in the C-CD denote the loci of MS and RGB stars, respectively.  Reddening vectors for 
$A_{\rm V} = 5$~mag and 10~mag are also drawn.  These diagrams can be
reliably utilized to distinguish the sources with infrared excess located to the right 
of the reddening vectors, away from the evolved MS and RGB stars.
\label{f:ngc346cmdccd}} 
\end{figure*} 
%%%%%%%%%%%%%%%%%%%%%%%%%%%%%%%%%%%%%%%%%%%%%%%%

%%%%%%%%%%%%%%%%%%%%%%%%%%%%%%%%%%%%%%%%%%%%%%%%
%\begin{figure}[t!]
%\centering 
%\includegraphics[width=0.975\columnwidth]{f11.eps} 
%\caption{$J-K_{\rm s}$ vs. $K_{\rm s}$ Color-Magnitude Diagram 
%of all sources found with our VLT/ISAAC photometry in the main part 
%of the star forming region NGC~346/N66 (small blue symbols), with
%the  263 candidate young stellar sources identified with our analysis
%overlaid (large red symbols). 
%\label{f:ysscmd}} 
%\end{figure} 
%%%%%%%%%%%%%%%%%%%%%%%%%%%%%%%%%%%%%%%%%%%%%%%%

\subsection{Selection of young stellar sources \label{s:ysoselect}}

To identify the sources that represent the most recent star 
formation in the region, we first consider the contribution of both 
the evolved MS {\sl and} RGB stellar populations of the SMC field,
the cluster BS~90, {\sl and} the association NGC~346 to the complete 
stellar sample. As shown in Figs.~\ref{f:totalcmds} and 
\ref{f:partscmds}, the MS is well defined with its brighter members
belonging to the association, but it is its red part and the region between 
the MS and RGB that host stellar sources at their earlier stages of 
formation \citep[see e.g.,][]{bik2006}. Therefore, and to eliminate 
the significant contamination of this part of the CMD by the
general field and BS~90, which is very close to the central part of N66, 
we constrain our analysis to the ISAAC FOV that covers the main body
of NGC~346/N66, defined as the central field with $-72.21 < {\rm Decl.}
({\rm deg}) \leq -72.15$ in the map of Fig.~\ref{f:xymap}. 
Color-composite images of 
this field, which covers the association NGC~346, the bar of N66, and 
several subclusters and {\sc Hii} regions in its vicinity, are shown in 
Fig.~\ref{f:colima}. The dust emission, seen in the 8~\micron\ band, 
indicates the centers of the most recent star formation and demonstrates 
the different information revealed by different wavelengths 
for the same region. 

The use of the near-IR C-CD is a reliable method for detecting 
sources, which are characterized by near-IR excess emission, 
from their positions in the diagram. In this diagram, the 
evolved populations are easily identified by comparison with 
models and considering interstellar extinction, PMS stars, namely 
classical and weak-line T~Tauri (CTT, WTT) 
stars \citep[e.g.,][]{appenzeller89}, Herbig Ae/Be (HAeBe) stars 
\citep[e.g.,][]{perez97,waters98}, and other YSOs \citep[e.g.,][]{lada84, 
andre94} are located in the reddest part. This is a signature of their
IR excess due to their circumstellar dust in the form of cocoons or 
disks. Hence, when selecting candidate YSOs and 
PMS stars, we consider only the sources that are detected in the
central ISAAC field in all three $JHK_{\rm s}$ wavebands. The 
$J-K_{\rm s}$, $K_{\rm s}$ CMD and $J-H$, $H-K_{\rm s}$ 
C-CD of these sources are shown in Fig.~\ref{f:ngc346cmdccd}. 
Considering that BS~90 to be a large cluster with a tidal radius between 
$r_{\rm t} \simeq$~2\farcm15 \citep{sabbi07} and 3\farcm15 \citep{rochau07}, 
its spatial extent covers the central field of  NGC~346/N66, and consequently 
affects the stellar content of the region. This is clearly demonstrated
in the CMD of Fig.~\ref{f:ngc346cmdccd}, where a prominent 
RGB can be seen. While these stars are bright in the near-IR, they
do not exhibit any significant excess emission. Our 
classification to identify young 
stellar sources based on their near-IR excess, therefore allows us to effectively 
discard most of the evolved stellar contaminants from our sample of selected 
young sources in formation. However, a definite selection requires 
a sophisticated multi-band study on an individual source basis, as we discuss in
Sect.~\ref{s:canysos}.

%%%%%%%%%%%%%%%%%%%%%%%%%%%%%%%%%%%%%%%%%%%%%%%%
\begin{figure*}[t!]
\centering 
\includegraphics[width=0.975\textwidth]{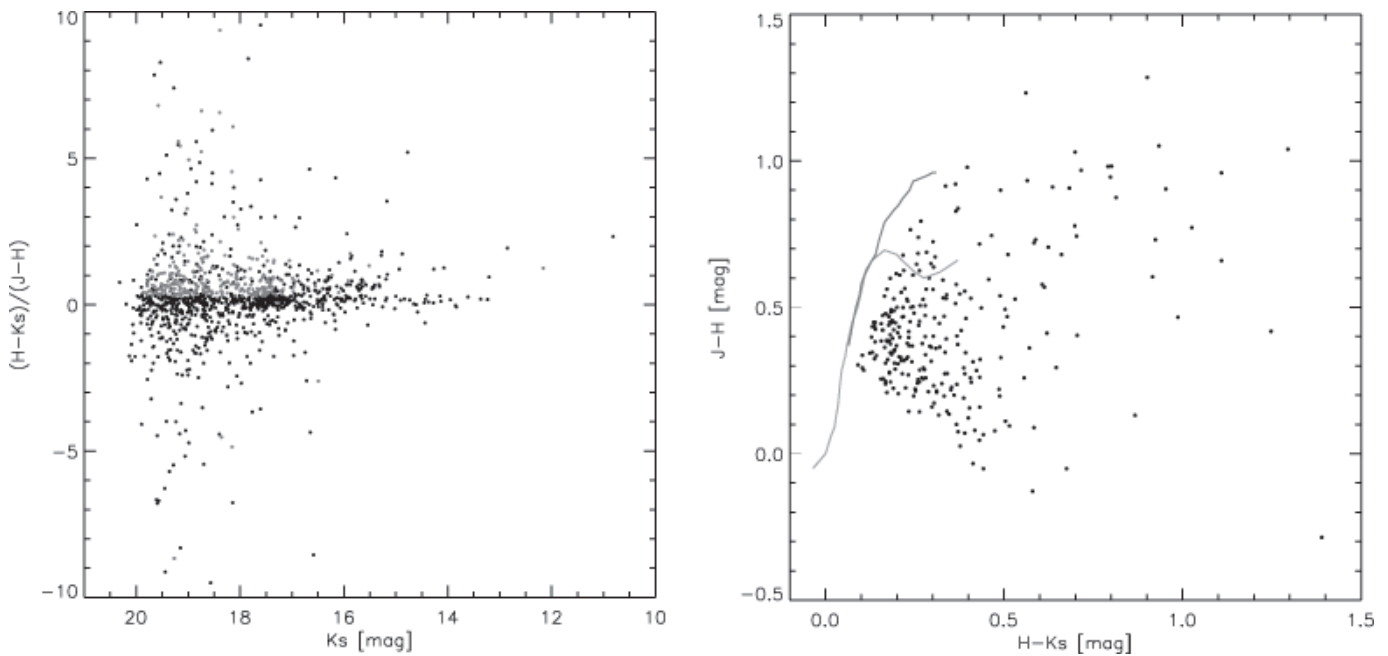} 
\caption{Selection of the sources identified to have a near-IR excess based on their
locations in the $H-K_{\rm s}$, $J-H$ C-CD, and $J-K_{\rm s}$, $K_{\rm s}$ CMD 
(see Fig.~\ref{f:ngc346cmdccd}). {\sl Left}: The fraction of
the color indices $(H-K_{\rm s})/(J-H)$ as a function of the $K_{\rm s}$
magnitudes of the detected sources.  {\sl Right}: The
C-CD diagram of all sources selected as the most prominent young stellar 
candidates with near-IR excess emission.\label{f:ccdredsel}} 
\end{figure*} 
%%%%%%%%%%%%%%%%%%%%%%%%%%%%%%%%%%%%%%%%%%%%%%%%

In the $J-K_{\rm s}$ vs. $K_{\rm s}$ CMD of Fig.~\ref{f:ngc346cmdccd}, the 
areas occupied by HAeBe \citep{eiroa02} and classical Be stars \citep{dougherty94} are 
delineated by two ellipses, and the positions of massive Galactic YSOs identified 
by \citet{hanson97} and \citet{bik2006} are indicated by green stellar symbols. 
In the C-CD of Fig.~\ref{f:ngc346cmdccd}, the typical positions of T~Tauri (TTS), 
HAeBe, and classical Be stars are drawn to indicate the loci, 
where these sources in our sample should be expected in this diagram. 
We note, however, that typical TTS are fainter than the detection limit of 
our photometry and therefore are barely discernible in the 
CMD of Fig.~\ref{f:ngc346cmdccd}, and only a few of them being seen in the C-CD. 
We refer, however, to these PMS stars for reasons of completeness. 
The diagrams of  Fig.~\ref{f:ngc346cmdccd} appear to comprise 
 a large variety of different stellar types at quite different evolutionary 
 stages. The regions in the CMD and C-CD, where classical Be stars
 are typically expected, also include MS populations. Searching for Be stars, 
 \citet{keller99} found in six fields centered on young clusters of the 
 Magellanic Clouds (NGC~346 included), that the average 
 fraction of Be to normal B stars is similar to that found in the Galaxy 
 \citep[$\sim$~20\%, see, e.g.,][p. 414]{cox00}. No connection between 
 the Be star fraction and age or metallicity was found by these authors, 
 and the classical Be stars detected in NGC~346 do not 
 have any influence on the evolution of the region. According to their near-IR 
 colors, red giants and subgiants are expected,  to 
 occupy the blue edges of the typical locations of HAeBe 
 stars. As a consequence, we should be careful to differentiate the sources that 
 are most probably at their earlier stages of evolution, based on their 
 near-IR excess, from the evolved red stars of the region.

Based on the discussion above, we make a first tentative selection
of the sources that most probably have near-IR excess using a diagram 
of the color fraction $(H-K_{\rm s})/(J-H)$ 
of all detected objects as a function of their brightness in 
$K_{\rm s}$. In this diagram, shown in Fig.~\ref{f:ccdredsel} 
(left), one can see that most sources are located along a horizontal 
sequence of stars with $(H-K_{\rm s})/(J-H) \simeq 0.0$, corresponding 
to the RGB and MS stars that exhibit no near-IR excess. Young stellar sources 
with a near-IR excess should be located away from the horizontal sequence.
We first assume that all sources with $(H-K_{\rm s})
 \lsim 0.0$  should not have any near-IR excess, as they are located 
 blueward of the MS in the C-CD of Fig.~\ref{f:ngc346cmdccd}.
We then apply a first-order selection of the remaining sources based on the criterion that 
an absolute offset from the horizontal sequence in the color fraction 
$(H-K_{\rm s})/(J-H)$ of about 0.3 is a reasonable limit to separate the
stars with near-IR excess from those that show no excess. As a consequence,
we select as sources with near-IR excess those that have color fractions 
$(H-K_{\rm s})/(J-H) \geq 0.3$ or $(H-K_{\rm s})/(J-H) \leq -0.3$ in the 
diagram of Fig.~\ref{f:ccdredsel} (left). 

% By default, we 
% consider as candidate young stellar sources only those with $(H-K_{\rm s}) > 0.0$ 
% (red part of the CMD). We then, tentatively
% select sources with  $(H-K_{\rm s})/(J-H) \geq 0.3$ and $(H-K_{\rm s})/(J-H) \leq -0.3$
% as those that demonstrate excess emission. We constrain further our selection 
% to sources that fall in the CMD at the positions expected for HAeBe stars and redder, 
% so that we eliminate the possibility
% of our sample being contaminated by MS and/or Classical Be stars.

However, an investigation of the positions of  these near-IR bright sources in the 
CMD of Fig.~\ref{f:ngc346cmdccd} shows that there is significant contamination of 
these objects by classical Be and RGB stars, which should be eliminated. 
 Therefore, we place yet tighter constraints on our sample by selecting only the 
 sources that fall at the positions 
expected for HAeBe stars as shown in Fig.~\ref{f:ngc346cmdccd} (left) and redder. 
More precisely, we select the sources that fulfill the following criteria: {\sl (i)} They are 
located to the blue part of the MS in the C-CD with $(H-K_{\rm s}) > 0.0$. {\sl (ii)} They 
have fractions of color indices $(H-K_{\rm s})/(J-H)$~$\geq 
+0.3$ or $\leq -0.3$. {\sl (iii)} They are located to the red part of a diagonal line that 
tangentially crosses the blue edge of the HAeBe area, specified in the $J-K_{\rm s}$,
$K_{\rm s}$ CMD of Fig.~\ref{f:ngc346cmdccd}. These sources 
represent our final sample of objects selected as the most likely star-forming candidates,
and the subject of our analysis here. The positions of these sources
are indicated by red  points in the diagram $(H-K_{\rm s})/(J-H)$ vs. $K_{\rm s}$ of Fig.~\ref{f:ccdredsel}. 
Their positions in the C-CD are also shown in Fig.~\ref{f:ccdredsel} 
(right panel). In this C-CD, it is indeed evident that our selected sample  
probably corresponds to objects with strong near-IR excess emission.

%%%%%%%%%%%%%%%%%%%%%%%%%%%%%%%%%%%%%%%%%%%%%%%%
\begin{figure*}[t!]
\centering 
\includegraphics[width=0.775\textwidth]{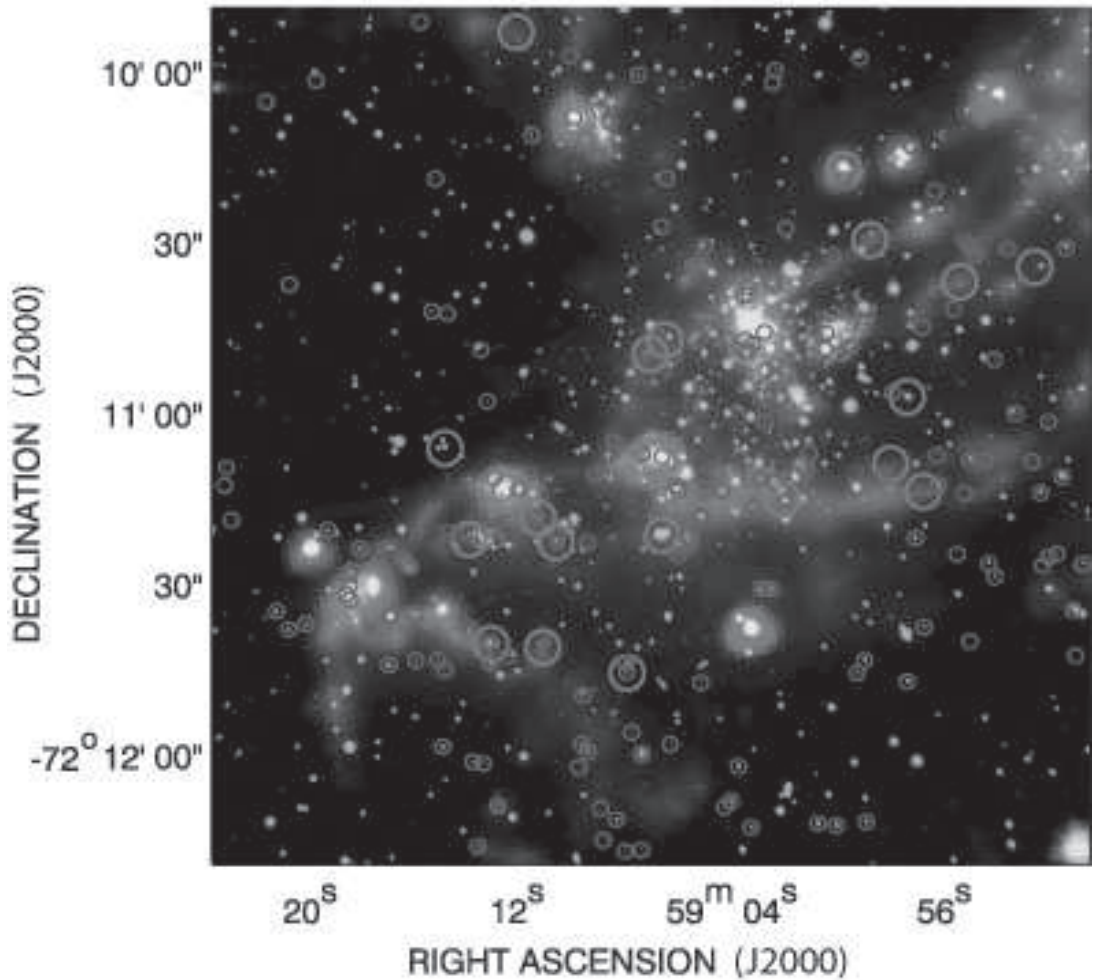} 
\caption{Positions of stellar sources in our photometric catalog, with  
 strong near-IR excess emission, based on their positions
in the $J-H$, $H-K_{\rm s}$ C-CD and $J-K_{\rm s}$, $K_{\rm s}$ CMD (see Fig.~\ref{f:ngc346cmdccd}). 
They are probably 
young stellar sources in the process of formation such as intermediate-mass PMS stars and 
high-mass YSOs. 
Their positions are overplotted with blue circle symbols on a color-composite
image consisting of the $K_{\rm s}$ ISAAC image (green) and {\sl Spitzer}/IRAC 
8~\micron\ filters (red).  Candidate YSOs from the catalog of \citet{simon07} are overplotted  with 
large red circular symbols. 
 \label{f:mapspitz}} 
\end{figure*} 
%%%%%%%%%%%%%%%%%%%%%%%%%%%%%%%%%%%%%%%%%%%%%%%%

%Most of the selected sources follow the filamentary dust emission
%seen in red at 8~\micron, suggesting that these objects are indeed related to
%ongoing star formation. However, since contamination by evolved stars is 
%expected, a more thorough investigation source by source is required for the
%final clarification of their nature.

\subsection{The sample of candidate young stellar sources in NGC~346/N66 \label{s:canysos}}

The selection scheme returned 263 candidate young stellar sources in the main
part of the region NGC~346/N66. The positions of these sources are 
shown in a map of this area in Fig.~\ref{f:mapspitz} as circular blue points. They are 
superimposed on 
a color-composite image constructed by combining our ISAAC image 
in the $K_{\rm s}$ filter (green component) with archived images 
obtained with the {\sl Infrared Array Camera} \citep[IRAC;][]{fazio04} 
onboard the {\sl Spitzer Space Telescope} and in particular channel 
4 (8.0~\micron, red component), within the GTO science program 63 
(PI: J. R. Houck). These IRAC data were used to detect candidate YSOs 
in the general region of N66 by \citet{simon07}.
The objects identified by these authors as ``definite'' YSOs are also
plotted in Fig.~\ref{f:mapspitz} with large red circular symbols.

In this map the positions of the detected 
candidate YSOs in NGC~346/N66 clearly correspond to dusty structures
seen in the 8~\micron\ emission as red filaments, these objects being 
located at
peaks of mid-IR emission in these filaments. These peaks can 
also be seen -- at lower spatial resolution -- in the
24~\micron\ band of {\sl Multiband Imaging Photometer for 
SIRTF} \citep[MIPS; e.g.][]{heim98, rieke03} onboard 
{\sl Spitzer}, and they are also observed in the 2.14~\micron\
H$_{\rm 2}$ line and the ISOCAM LW2 band that covers 
5~$-$~8~\micron\ \citep{contursi00,rubio00}. We note that 
the spatial distribution of our candidate young stellar sources 
follows the filamentary dust emission,
being concentrated along a few southern dusty arcs (at the middle 
and bottom of the map) and one northern dusty arm (at the
top of the image). The latter, along with the south-eastern
filament, is understood to be the product of a more recent
triggered star formation event \citep{gouliermis08}. 
\citet{hennekemper08} discuss the spatial 
distribution of the low-mass PMS stars detected by 
\citet{gouliermis06}, and of sources with 
excess H\alp\ emission that are the most likely candidates 
to be intermediate-mass PMS stars (their Figs. 1 and 2). 
The spatial distribution of all these objects follows 
the same trend as our sources, and it is probable that  
both catalogs include several objects in common. In addition, the positions
of a significant number of candidate young stellar sources found in 
our near-IR photometry, shown in Fig.~\ref{f:mapspitz}, also seem 
to be quite clustered in several concentrations. We also note that almost all of these 
concentrations coincide with subclusters of PMS stars observed 
with HST/ACS identified by applying two cluster 
analysis techniques \citep{schmeja09}. 

We  first compared a nominal search-box of 2\arcsec\ 
of the catalog of our sources with that of the YSOs found 
with {\sl Spitzer} by \citet{simon07} and PMS stars with H\alp\ 
excess found with {\sl Hubble} by \citet{hennekemper08} in the 
same region. This comparison returns more than $\sim$~40\% of 
the H\alp\ excess stars and $\sim$~60\% of the YSOs also detected in 
the near-IR by ourselves. There are cases where more than one objects 
found with {\sl Hubble} coincide with one of our sources, while 
in other cases a few of our sources are found as components of one 
of the YSOs found with {\sl Spitzer} within the specified search-box. 
This phenomenon is caused by differences in the 
resolving efficiency of the three instruments, demonstrating 
the importance of resolving multiple sources into their 
true components and classifying their true nature. The resolution 
achieved by VLT/ISAAC is at least 10 times higher than that of 
{\sl Spitzer}/IRAC, allowing a reasonable identification of any components
in multiple YSOs. However, 15\% of our sources coinciding  with 
H{\alp}-excess objects identified by \citet{hennekemper08},
are resolved by HST/ACS to be multiple systems {\sl themselves}. 
Therefore, the differences in resolution between the 
data sets obtained with HST, VLT, and {\sl Spitzer} is a crucial issue 
in identifying {\sl true} single-objects, or the components 
of multiple systems in our sample. As we discuss later, to perform 
a more detailed study in the near-IR, observations of the highest 
possible spatial resolution are essential.

%The low percentage of YSOs 
%detected with {\sl Spitzer} also found in our sample demonstrates that 
%possibly the remaining objects may be at a very early 
%stage of their formation, and therefore identified only in mid- and far-IR 
%wavelengths. 

It is almost certain that our sample of candidate PMS stars and YSOs 
is incomplete, since for example TTS are not included 
because of our detection limit. Other young stellar sources such as HAeBe
stars may also be missing due to the strong constraints of our selection criteria. 
In addition, our sample may be contaminated by 
evolved stars, which in general do not exhibit any significant 
near-IR excess. While the positions of most of the sources in our catalog coincide 
with the dust filaments of N66, illustrating their youthfulness, there are 
candidate young sources in our sample that are located away from 
the dusty filaments of Fig.~\ref{f:mapspitz}. These are usually assumed to be
evolved stars, or even background galaxies, rather than stellar sources in formation. 
However, decontaminating our sample of old stars in the field and BS~90
located in the central region of NGC~346/N66 
on a statistical basis for such a small sample, or based on their 
positions away from the dust filaments, will possibly compromise the catalog 
of sources with true near-IR excess and it produce selection effects. 

A more sophisticated selection on a source-by-source basis 
is certainly required to identify
the most prominent objects that represent the most recent star 
formation in the region. This, however, can only be achieved
through the excessive use of imaging in many different wavebands
and/or spectroscopy \citep[e.g.,][]{chen09}, so that complete SEDs 
of individual sources can be constructed and consequently their true 
nature accurately defined \citep[e.g.,][]{whitney08}. Observations of 
higher spatial resolution will allow the components of any unresolved bright sources 
in our sample to be recognized. Such a thorough analysis would certainly 
include the use of previous observations of the region NGC~346/N66 
from various telescopes at different wavelengths including those 
of {\sl HST} \citep{gouliermis06, hennekemper08} and {\sl Spitzer} 
\citep{bolatto07, simon07}, as well as {\sl new} near-IR observations
of higher spatial resolution and sensitivity. A preliminary study 
of the available data and the preparation of follow-up observations 
in the near-IR is currently being performed by ourselves.

\section{Conclusions}

We have presented a detailed near-IR photometric study with VLT/ISAAC
of the star-forming region NGC~346/N66 in the SMC. We have used
archival ISAAC imaging of the general area of N66, which
includes the stellar association NGC~346, the intermediate-age 
cluster BS~90, and a southern empty control field of the SMC. We 
have performed photometry on images obtained in the filters $J$ (1.25~\micron), 
$H$ (1.65~\micron), and $K_{\rm s}$ (2.16~\micron) and derived a catalog
of more than 2~500 stars detected in all three wavebands. The 
color-magnitude diagrams of these stars include a collection of
different stellar populations, comprising the young MS stars
of the association mixed with the RGB and old MS stars of BS~90
and the general field of the SMC, but also objects that are at their
earlier stages of their formation. We select the best PMS and YSO 
candidates in our sample on the basis of their positions in the 
near-IR color-color and color-magnitude diagrams.

We focus the selection of these sources on the central field
observed with ISAAC, which covers only the main part of the
nebula N66 and the association NGC~346  to avoid 
any severe contamination of our sample with the evolved red
stars of BS~90 and the field. In this area, our photometry 
detected 1~174 stars in all three wavebands for which
the near-IR CMD and C-CD are constructed.
In these diagrams, the evolved stellar populations are
mostly aligned along the sequences of RGB and MS stars as they 
are expected to be by the evolutionary models depending on the  
interstellar extinction, but certainly contaminate the sample of young 
stellar sources.  The reason is that, while these  
sources, such as low-mass T~Tauri stars, intermediate-mass 
Herbig Ae/Be stars, and massive YSOs, exhibit excess emission 
in the near-IR due to circumstellar dust, their positions in the 
CMD and C-CD do not necessarily cover the reddest part.

Bearing this in mind, we make a tentative selection of PMS and YSO 
candidates with $(H-K_{\rm s}) \gsim~0.0$ (redder than MS and RGB) 
in the diagram of the color fraction $(H-K_{\rm s})/(J-H)$ as a function 
of the $K_{\rm s}$ brightness. In this diagram, sources that have  
an excess in their near-IR colors are located away from the horizontal 
sequence of evolved stars. We select the sources that have an offset 
from the horizontal sequence of $|\Delta(H-K_{\rm s}/J-H)| \geq 0.3$ 
and we decontaminate the sample by selecting sources located in the 
area of the $J-K_{\rm s}$, $K_{\rm s}$ CMD that HAeBe stars are expected 
to occupy and with redder colors. We consider 
the selected objects as the most probable candidates
to be stars in formation. This selection delivers 263 candidate
young stellar sources, which are located along the dusty filamentary 
structures of N66 seen in the 8~\micron\ emission from {\sl Spitzer}
and the 2.14~\micron\ H$_{\rm 2}$ line. 

Combining observations at several wavelengths to construct complete SEDs 
of individual sources is the most accurate means of establishing their true 
nature. Objects from our catalog of young stellar sources  do indeed coincide 
with candidate YSOs detected with {\sl Spitzer} and sources with excess emission 
in H\alp\ in the region observed with {\sl Hubble}. However, since a large amount 
of data per object is required for detailed SED studies, it is necessary to enhance 
the available data sets with new data, preferably obtained with cameras of higher 
resolving power, so that multiple objects can be resolved in their components. 

\begin{acknowledgements}

We thank M. Rubio and R. H. Barb\'a for their comments 
and suggestions. D. A. G. kindly acknowledges the support 
by the Deutsche Forschungsgemeinschaft (DFG) through grant 
GO~1659/1-2. This research has made use of the SIMBAD 
database, operated at the CDS, Strasbourg, France, of 
NASA's Astrophysics Data System, and images obtained 
with the {\sl Spitzer Space Telescope}, which is operated by 
the Jet Propulsion Laboratory, California Institute of 
Technology under a contract with NASA. It is also based on 
observations made with ESO Telescopes at the La Silla 
Paranal Observatory under program ID 063.I-0329. 

\end{acknowledgements}

\bibliographystyle{aa}
\bibliography{12226_ms}

\end{document}